# THE DIRAC FIELD IN REAL DOMAIN

Krunoslav Ljolje
Academy of Sciences and Arts of Bosnia and Herzegovina,
Sarajevo,Bosnia and Herzegovina

Abstract

The Dirac field is analysed in real domain.A connection with the classical field theory is shown and corresponding relation to the quantum physics.

## 1. INTRODUCTION

The Dirac field has appeared in the past as a part of procedure called "quantization",substitution of physical quantities by operators and associated aspects.In the variational language the Dirac field function has been considered as a Lagrange's variable.In the article [1] and later on in [2-6] this interpretation was called in question and canonical character of this field function has been established. Analogy with the classical electromagnetic field has been demonstrated.Complex description was retained,also in the case of interaction of the Dirac field with the electromagnetic field where the substitution $i\partial_\alpha \rightarrow i\partial_\alpha - eA_\alpha$ has been applied.

The analogy with the classical electromagnetic field suggests full description of the Dirac field in real domain [7].We consider this point in this article.

The free field is analysed in Section 2.Interaction of the Dirac field with the electromagnetic field in real domain is presented in Section 3.By this the basic elements of the Dirac field have been put in the framework of the classical field theory.

The problem of the self-interaction is not analysed in this article.

This analysis has the following values : (1) clear physical interpretation of the field functions and corresponding processes, (2) adequate mathematical description and (3) a new look at the background of the quantum physics.

## 2. FREE FIELD

Equations of the free Maxwell(-Lorentz) field

$$\partial_\alpha F^{\alpha\beta} = 0 \quad , \quad \partial_\alpha \widetilde{F}^{\alpha\beta} = 0 \quad , \tag{1}$$

where

$$F^{\alpha\beta} = \begin{bmatrix} 0 & -E_x & -E_y & -E_z \\ E_x & 0 & -B_z & B_y \\ E_y & B_z & 0 & -B_x \\ E_z & -B_y & B_x & 0 \end{bmatrix} \quad , \quad \widetilde{F}^{\alpha\beta} = \frac{1}{2}\varepsilon^{\alpha\beta\xi\zeta}F_{\xi\zeta} \quad , \tag{2}$$

can be written in the form [5-6]

$$D\Psi = 0 , \tag{3}$$

where

$$D = \partial_\alpha \eta^\alpha , \tag{4}$$

$$\eta^o = \begin{bmatrix} 1 & 0 \\ 0 & -1 \end{bmatrix}_{8x8} \quad , \quad \eta^i = \begin{bmatrix} 0 & a^i \\ -a^{i+} & 0 \end{bmatrix} \quad ,$$

$$a^1 = \begin{bmatrix} 0 & 0 & 0 & 1 \\ 0 & 0 & 1 & 0 \\ 0 & -1 & 0 & 0 \\ 1 & 0 & 0 & 0 \end{bmatrix}, \; a^2 = \begin{bmatrix} 0 & 0 & -1 & 0 \\ 0 & 0 & 0 & 1 \\ 1 & 0 & 0 & 0 \\ 0 & 1 & 0 & 0 \end{bmatrix}, \; a^3 = \begin{bmatrix} 0 & 1 & 0 & 0 \\ -1 & 0 & 0 & 0 \\ 0 & 0 & 0 & 1 \\ 0 & 0 & 1 & 0 \end{bmatrix},$$

$$\eta^\alpha \eta^\beta + \eta^\beta \eta^\alpha = 2g^{\alpha\beta} \; , \tag{6}$$

$$\Psi = \begin{bmatrix} E_x \\ E_y \\ E_z \\ F \\ B_x \\ B_y \\ B_z \\ G \end{bmatrix} \quad , \qquad \text{F,G are scalars.}$$

(7)

All quantities in Eqs. (1-7) are real.

The Lagrangian density of the field $\Psi$ is given by

$$L = K\overline{\Psi}\Psi \quad , \tag{8}$$

where

$$\Psi = D\Phi \quad , \quad \overline{\Psi} = \Psi^+ \eta^o \quad , \tag{9}$$

$$\Phi = \begin{bmatrix} -A_x \\ -A_y \\ -A_z \\ f \\ C_x \\ C_y \\ C_z \\ -\varphi \end{bmatrix} . \tag{10}$$

K is constant (for the electromagnetic field equal to $1/8\pi$).

Lagrange's variables are components of $\Phi$ .For convenience of calculation we may take $\Phi$ to be complex and then consider only real solutions (but it is not necessary).In that case $(\Phi, \Phi^+)$ are the Lagrange's variables.

According to

$$\frac{\partial L}{\partial u} - \partial_\mu \frac{\partial L}{\partial(\partial_\mu u)} = 0 \tag{11}$$

the Lagrange's equations are given by

$$\partial_\alpha \partial^\alpha \Phi = 0 \quad , \quad \partial_\alpha \partial^\alpha \Phi^+ = 0 \ . \tag{12}$$

The conjugate momenta to $\Phi, \Phi^+$ are

$$\begin{aligned} c\Pi_\Phi &= \frac{\partial L}{\partial(\partial_o \Phi)} = K\psi^+, \\ c\Pi_{\Phi^+} &= \frac{\partial L}{\partial(\partial_o \Phi^+)} = K\psi. \end{aligned} \tag{13}$$

Corresponding Hamiltonian and canonical equations are given by

$$\begin{aligned} H &= c\Pi_\Phi \partial_o \Phi + c\partial_o \Phi^+ \Pi_{\Phi^+} - L = \\ &= \frac{c^2}{K} \Pi_\Phi \eta^o \Pi_{\Phi^+} - c\Pi_\Phi \eta^o \partial_j \eta^j \Phi - c(\partial_j \eta^j \Phi)^+ \eta^o \Pi_{\Phi^+}, \end{aligned} \tag{14}$$

$$\begin{aligned} c\partial_o \Phi &= \frac{\partial H}{\partial \Pi_\Phi} = \frac{c^2}{K} \eta^o \Pi_{\Phi^+} - \eta^o c\partial_j \eta^j \Phi \longrightarrow \frac{c}{K} \Pi_{\Phi^+} = \psi, \\ c\partial_o \Phi^+ &= \frac{\partial H}{\partial \Pi_{\Phi^+}} = \frac{c^2}{K} \Pi_\Phi \eta^o - c(\partial_j \eta^j \Phi)^+ \eta^o \longrightarrow \frac{c}{K} \Pi_\Phi = \psi^+, \end{aligned} \tag{15}$$

$$\begin{aligned} c\partial_o \Pi_\Phi &= -\frac{\partial H}{\partial \Phi} + \partial_j \frac{\partial H}{\partial(\partial_j \Phi)} = -c\partial_j \Pi_\Phi \eta^o \eta^j \longrightarrow (D\psi)^+ = 0, \\ c\partial_o \Pi_{\Phi^+} &= -\frac{\partial H}{\partial \Phi^+} + \partial_j \frac{\partial H}{\partial(\partial_j \Phi^+)} = -c\partial_j \eta^{j+} \eta^o \Pi_{\Phi^+} \longrightarrow D\psi = 0. \end{aligned} \tag{16}$$

Written in terms of the components of $\Phi$ and $\Psi$ they become

$$\begin{aligned} \partial_\alpha (F^{\alpha\beta} + g^{\alpha\beta} G) &= 0, \\ \partial_\alpha (\tilde{F}^{\alpha\beta} + g^{\alpha\beta} F) &= 0, \end{aligned} \tag{17}$$

$$\begin{aligned} F^{\alpha\beta} &= \partial^\alpha A^\beta - \partial^\beta A^\alpha - \frac{1}{2} \varepsilon^{\alpha\beta\xi\zeta} (\partial_\xi C^\zeta - \partial_\zeta C^\xi), \\ \tilde{F}^{\alpha\beta} &= \partial^\alpha C^\beta - \partial^\beta C^\alpha + \frac{1}{2} \varepsilon^{\alpha\beta\xi\zeta} (\partial_\xi A^\zeta - \partial_\zeta A^\xi), \\ F &= \partial_\alpha C^\alpha, \quad G = \partial_\alpha A^\alpha. \end{aligned} \tag{18}$$

Taking F=G=0 (they are scalars!) and $C^\alpha = 0$, equations (16-17) become equations of the standard classical free electromagnetic field.

Inspection of equations (3,4,6-10) shows that $(\Phi,\Psi)$ may be also a spinor field [5] (composed by means of spinors). Consequently, starting with L given by (8) equations (15-16) are also canonical equations of a spinor massless field.

An extension to a spinor field with a "mass" term is given by

$$L = K\left(\overline{\psi}\psi - \kappa^2\overline{\Phi}\Phi\right), \tag{19}$$

where κ is a real constant(positive or negative).

Corresponding Lagrange′s and canonical equations are given by

$$\left(\partial_\alpha\partial^\alpha + \kappa^2\right)\Phi = 0, \quad \left(\partial_\alpha\partial^\alpha + \kappa^2\right)\Phi^+ = 0, \tag{20}$$

$$\begin{aligned} &\partial_o\Phi = \eta^o\psi - \partial_j\eta^o\eta^j\Phi \quad \longrightarrow \psi = D\Phi, \\ &\partial_o\Phi^+ = \psi^+\eta^o - \partial_j\Phi^+\eta^o\eta^j \quad \longrightarrow \psi^+ = (D\Phi)^+, \end{aligned} \tag{21}$$

$$\begin{aligned} &c\partial_o\Pi_\Phi = -K\kappa^2\overline{\Phi} - K\partial_j\psi^+\eta^o\eta^j \quad \longrightarrow \left(\overline{D\psi}\right) + \kappa^2\overline{\Phi} = 0, \\ &c\partial_o\Pi_{\Phi^+} = -K\kappa^2\eta^o\Phi - K\partial_j\eta^o\eta^j\psi \quad \longrightarrow D\psi + \kappa^2\Phi = 0. \end{aligned} \tag{22}$$

For the spinor field the following linear combination of Φ and ψ (they are spinors!) are possible and instructive

$$\begin{aligned} \psi_I &= \kappa\Phi + i\psi, \\ \psi_{II} &= \kappa\Phi - i\psi, \end{aligned} \tag{23}$$

From here one gets

$$\begin{aligned} \Phi &= \frac{1}{2\kappa}\left(\psi_I + \psi_{II}\right), \\ \psi &= \frac{1}{2i}\left(\psi_I - \psi_{II}\right). \end{aligned} \tag{24}$$

Equations (21-22) then become

$$\begin{aligned} &\left(i\partial_\alpha\eta^\alpha - \kappa\right)\psi_I = 0, \\ &\left(i\partial_\alpha\eta^\alpha + \kappa\right)\psi_{II} = 0 \qquad \text{and a.e.} \end{aligned} \tag{25}$$

These are Dirac equations with eight-component functions and with positive and negative mass terms. Both equations describe the field. For real field it follows

$$\psi_{II} = \psi_I{}^*. \tag{26}$$

The unitary matrix [7]

$$
S=\frac{1}{\sqrt{2}}\begin{bmatrix} -1 & i & 0 & 0 & 0 & 0 & 0 & 0 \\ 0 & 0 & 1 & -i & 0 & 0 & 0 & 0 \\ 0 & 0 & 0 & 0 & -i & -1 & 0 & 0 \\ 0 & 0 & 0 & 0 & 0 & 0 & i & -1 \\ 0 & 0 & -1 & -i & 0 & 0 & 0 & 0 \\ -1 & -i & 0 & 0 & 0 & 0 & 0 & 0 \\ 0 & 0 & 0 & 0 & 0 & 0 & -i & -1 \\ 0 & 0 & 0 & 0 & -i & 1 & 0 & 0 \end{bmatrix}, \quad S^{-1}=S^{+}, \tag{27}
$$

transforms Eqs. (25) into

$$
\left(i\partial_\alpha\gamma^\alpha-\kappa\right)\left(S\Psi_I\right)_a=0 \ , \ \left(i\partial_\alpha\gamma^\alpha-\kappa\right)\left(S\Psi_{II}\right)_b=0, \tag{28}
$$

where $\gamma^\alpha$ are standard Dirac matrices.

There is a family of solutions given by

$$
\Psi=\kappa N\Phi \ , \tag{29}
$$

$$
\left(D-\kappa N\right)\Phi=0 \ , \tag{30}
$$

where

$$
N=\begin{bmatrix} N_a & 0 \\ 0 & N_b \end{bmatrix} \ , \ N_a=\begin{bmatrix} 0 & -1 & 0 & 0 \\ 1 & 0 & 0 & 0 \\ 0 & 0 & 0 & -1 \\ 0 & 0 & 1 & 0 \end{bmatrix} \ , \ N_b=\begin{bmatrix} 0 & -1 & 0 & 0 \\ 1 & 0 & 0 & 0 \\ 0 & 0 & 0 & 1 \\ 0 & 0 & -1 & 0 \end{bmatrix} \ , \tag{31}
$$

$$
N^{+}=-N \ , \ N^2=-1 \ , \ N\eta^\alpha=\eta^\alpha N \ . \tag{32}
$$

Indeed,after substitution of (29) into (23) one gets

$$
\Psi_I=\kappa\left(1+iN\right)\Phi \ , \ \Psi_{II}=\kappa\left(1-iN\right)\Phi \tag{33}
$$

and then from (25) it follows (30).

Application of S-transformation (27) to (30) yields

$$
S\Phi=\Phi'=\begin{bmatrix} \Phi_a' \\ \Phi_b' \end{bmatrix} \ , \tag{34}
$$

$$
\left(i\partial_\alpha\gamma^\alpha-\kappa\right)\Phi_a'=0 \ , \ \left(i\partial_\alpha\gamma^\alpha+\kappa\right)\Phi_b'=0 \ , \tag{35}
$$

where $\Phi_a', \Phi_b'$ are four-component matrices.

The family (29-30) is interesting since the field $\Phi$ itself,written in the complex form (34),satisfies the Dirac equations (35).

Constants of motion .

The scalar constant of motion comes from

$$\partial_\alpha\left(\overline{\Psi}_I\eta^\alpha\Psi_I\right)=0 \tag{36}$$

and is given by

$$Q=\text{const.}_Q\int\Psi_I^+\Psi_I d^3x=\text{const.}_Q\int\left(\kappa^2\Phi^+\Phi+\Psi^+\Psi\right)d^3x\ . \tag{37}$$

For the solutions (29-30) it is

$$Q=\text{const.}_Q\int 2\kappa^2\Phi^+\Phi d^3x\ . \tag{38}$$

The energy-momentum constant of motion and the spin projection $S^3$ constant of motion are given by

$$P^\alpha=\text{const.}_P K\int\left(\Psi^+\partial^\alpha\Phi-\Phi^+\partial^\alpha\Psi\right)d^3x\ , \tag{39}$$

$$S^3=\text{const.}_M\frac{K}{2}\int\left(\Psi^+\eta^1\eta^2\Phi-\Phi^+\eta^{1+}\eta^{2+}\Psi\right)d^3x\ . \tag{40}$$

For the solutions (29-30) they are

$$P^\alpha=\text{const.}_P(-2\kappa K)\int\Phi^+\partial^\alpha N\Phi d^3x\ , \tag{41}$$

$$S^3=\text{const.}_M(\kappa K)\int\Phi^+\eta^1\eta^2 N\Phi d^3x\ . \tag{42}$$

Example

At the first let us mention that from (35) it follows

$$\Phi_b'=N_b\Phi_a'^*\ . \tag{43}$$

Here we have made use of

$$N_b\gamma^\beta=\gamma^\beta N_b\quad(\beta\neq 2)\ ,\quad N_b\gamma^2=-\gamma^2 N_b\ . \tag{44}$$

Thus,one gets

$$\Phi'=\begin{bmatrix}\Phi_a'\\ N_b\Phi_a'^*\end{bmatrix}\ . \tag{45}$$

Let us consider the stationary solution

$$\Phi_a'=\sqrt{\frac{k_o+\kappa}{2k_o}}\begin{bmatrix}1\\0\\ \dfrac{k}{k_o+\kappa}\\0\end{bmatrix}e^{-i(k_o x^o-kz)}\ ,\quad k_o=\sqrt{\kappa^2+k^2}\ . \tag{46}$$

Substitution in (45) yields

$$\Phi' = \sqrt{\frac{k_o+\kappa}{2k_o}} \begin{bmatrix} \begin{bmatrix} 1 \\ 0 \\ \dfrac{k}{k_o+\kappa} \\ 0 \end{bmatrix} e^{-i(k_o x^o - kz)} \\ \begin{bmatrix} 0 \\ 1 \\ 0 \\ -\dfrac{k}{k_o+\kappa} \end{bmatrix} e^{i(k_o x^o - kz)} \end{bmatrix} \tag{47}$$

From here it follows

$$\Phi = \sqrt{\frac{k_o+\kappa}{k_o}} \begin{bmatrix} -\cos(k_o x^o - kz) \\ -\sin(k_o x^o - kz) \\ 0 \\ 0 \\ \dfrac{k}{k_o+\kappa}\sin(k_o x^o - kz) \\ -\dfrac{k}{k_o+\kappa}\cos(k_o x^o - kz) \\ 0 \\ 0 \end{bmatrix} \tag{48}$$

The constants of motion (38,41 and 42) for this solution are given by

$$Q = \mathrm{const.}_Q 4\kappa^2 L^3 \quad , \tag{49}$$

$$P^\alpha = \mathrm{const.}_P 4\kappa^2 L^3 \left(\frac{K}{\kappa} k^\alpha\right) \quad , \tag{50}$$

$$S^3 = \mathrm{const.}_M 4\kappa^2 L^3 \left(\frac{K}{2\kappa}\right) \quad . \tag{51}$$

Selection

$$\mathrm{const.}_Q 4\kappa^2 L^3 = 1 \quad , \quad \mathrm{const.}_P 4\kappa^2 L^3 = \frac{1}{c} \quad , \quad \mathrm{const.}_M 4\kappa^2 L^3 = \frac{1}{c} \quad , \tag{52}$$

yields

$$\begin{aligned} Q &= 1, \\ P^\alpha &= \hbar k^\alpha, \\ S^3 &= \frac{1}{2}\hbar. \end{aligned} \tag{53}$$

This may be interpreted as a state of a “particle” with the energy-momentum $\hbar k^\alpha$ and “spin state” $1/2$.

Let us notice that in the function $\Phi$ is present not only Dirac solution (46) but also $\Phi_b{}'$ which corresponds to the Dirac negative energy state with oposite momentum and spin $-\frac{1}{2}$.

## 3.INTERACTION WITH THE ELECTROMAGNETIC FIELD

In this part we consider interaction of the real Dirac field with the (real) electromagnetic field.

We start with

$$L_D \longrightarrow L_{DI} = K\left[\overline{\Psi}F_1(A^\alpha)\Psi - \kappa^2\overline{\Phi}F_2(A^\alpha)\Phi\right] \quad , \tag{54}$$

where the functions $F_1$ and $F_2$ are subjected to condition that the equation for the real Dirac field becomes equation that is identical in form to the standard Dirac equation with the electromagnetic field.

Variation of S with respect to $\overline{\Phi}$ yields

$$\delta_{\overline{\Phi}}S = \int \delta\overline{\Phi}\left[-D(F_1\Psi) - \kappa^2 F_2\Phi\right]d^4x = 0$$

and

$$D(F_1\Psi) + \kappa^2 F_2\Phi = 0 \quad . \tag{55}$$

Introduction of the new functions $\Psi_I, \Psi_{II}$ according to

$$\begin{aligned} F_1\Psi &= \frac{1}{2i}(\Psi_I - \Psi_{II}), \\ \kappa\Phi &= \frac{1}{2}(\Psi_I + \Psi_{II}), \end{aligned} \tag{56}$$

in the equation (55) and $\Psi = D\Phi$ yields

$$\left[iD - \frac{\kappa}{2}(F_2 + F_1^{-1})\right]\Psi_I - \frac{\kappa}{2}(F_2 - F_1^{-1})\Psi_{II} = 0 \quad , \tag{57}$$

$$\left[iD + \frac{\kappa}{2}(F_2 + F_1^{-1})\right]\Psi_{II} + \frac{\kappa}{2}(F_2 - F_1^{-1})\Psi = 0 \quad . \tag{58}$$

From here it follows

$$F_2 - F_1^{-1} = 0 \quad , \tag{59}$$

$$\frac{1}{2}(F_2 + F_1^{-1}) = 1 + \frac{e}{K}A_\beta\eta^\beta \quad . \tag{60}$$

Solutions of these equations are given by

$$F_2 = 1 + \frac{e}{K} A_\beta \eta^\beta \quad , \tag{61}$$

$$F_1 = \frac{1 - \frac{e}{K} A_\beta \eta^\beta}{1 - \frac{e^2}{K^2} A_\beta A^\beta} \quad . \tag{62}$$

After substitution of $F_1$ and $F_2$ into (54) one gets

$$\mathsf{L}_{DI} = K\left[ \frac{\overline{\Psi}(1-a)\Psi}{1-a^2} - \kappa^2 \overline{\Phi}(1+a)\Phi \right] \quad , \tag{63}$$

where

$$a = \frac{e}{K} A_\beta \eta^\beta \quad . \tag{64}$$

Therefore, the Lagrangian density of the real Dirac field interacting with (real) electromagnetic field is given by

$$\mathsf{L} = \mathsf{L}_{em} + \mathsf{L}_{DI} + (\mathsf{L}_{ext}) \quad . \tag{65}$$

The linear approximation of $\mathsf{L}_{DI}$ with respect to $A^\alpha$ is given by

$$\mathsf{L} \approx \mathsf{L}_D - \frac{1}{c} A_\alpha j^\alpha_{Do} \tag{66}$$

where

$$\frac{1}{c} j^\alpha_{Do} = e\left( \overline{\Psi}\eta^\alpha \Psi + \kappa^2 \overline{\Phi}\eta^\alpha \Phi \right) \tag{67}$$

in accordance with the classical (linear) electrodynamics.

Returning to (63) the conjugate momenta to $\Phi$ and $\Phi^+$ are given by

$$\Pi_\Phi = \frac{K}{c} \frac{\overline{\Psi}(1-a)\eta^o}{1-a^2} \quad , \tag{68}$$

$$\Pi_{\Phi^+} = \frac{K}{c} \frac{(1-a)\Psi}{1-a^2} \tag{69}$$

Corresponding Hamiltonian density and canonical equations are given by, respectively,

$$\mathsf{H}_{DI} = \frac{e^2}{K} \Pi_\Phi \eta^o (1+a) \Pi_{\Phi^+} - c\Pi_\Phi \partial_j \eta^o \eta^j \Phi - e\partial_j \overline{\Phi} \eta^j \Pi_{\Phi^+} +$$

$$+ K\kappa^2 \overline{\Phi}(1+a)\Phi \quad , \tag{70}$$

$$D\Phi - (1+a)\frac{c}{K} \Pi_{\Phi^+} = 0 \quad , \tag{71}$$

$$\overline{D\Phi} - \frac{c}{K}\Pi_{\Phi}\eta^{o}(1+a) = 0 \quad , \tag{72}$$

$$D\left(\frac{c}{K}\Pi_{\Phi^+}\right) + \kappa^2(1+a)\Phi = 0 \quad , \tag{73}$$

$$\overline{D\left(\frac{c}{K}\Pi_{\Phi^+}\right)} + \kappa^2\overline{\Phi}(1+a) = 0 \quad . \tag{74}$$

According to (56),(61) and (62) these equations may be written in the form

$$\frac{1}{\kappa}D(\Psi_I + \Psi_{II}) + i(1+a)(\Psi_I - \Psi_{II}) = 0 \quad , \tag{75}$$

$$\frac{1}{\kappa}D(\Psi_I - \Psi_{II}) + i(1+a)(\Psi_I + \Psi_{II}) = 0 \quad , \tag{76}$$

and a.e.From here one gets

$$[iD - \kappa(1+a)]\Psi_I = 0 \quad , \tag{77}$$

$$[iD + \kappa(1+a)]\Psi_{II} = 0 \quad , \tag{78}$$

or

$$[iD - \kappa(1+a)]\Psi_I = 0 \quad , \quad \Psi_{II} = \Psi_I^* \quad . \tag{79}$$

The equation (77) is standard Dirac equation with eight-component field function.The unitary matrix S transforms it into four-component Dirac functions

$$[i\partial_\alpha\gamma^\alpha - \kappa(1+a)](S\Psi_I)_a = 0 \quad , \tag{80}$$

$$[i\partial_\alpha\gamma^\alpha - \kappa(1+a)](S\Psi_I)_b = 0 \quad , \tag{81}$$

where

$$\Psi_I' = S\Psi_I \equiv \begin{bmatrix}(S\Psi_I)_a\\(S\Psi_I)_b\end{bmatrix} = \begin{bmatrix}\Psi_{Ia}'\\\Psi_{Ib}'\end{bmatrix} \quad . \tag{82}$$

$(S\Psi_I)_a$ and $(S\Psi_I)_b$ are four-component matrices.It is important to notice that $\Psi_I$ is composed of the Dirac field function $\Phi$ and the electromagnetic field functions.

Similarly to (29-30) there is a family of solutions

$$\Psi = (1+a)\kappa N\Phi \quad , \tag{83}$$

$$[D - \kappa(1+a)N]\Phi = 0 \quad . \tag{84}$$

In the next we restrict attention to this family of solutions.

S-matrix transforms Eq. (84) into

$$[i\partial_\alpha\gamma^\alpha - \kappa(1+\tilde{a})]\varphi_a = 0 \quad , \quad \tilde{a} = \frac{e}{K}A_\beta\gamma^\beta \quad , \tag{85}$$

$$\left[i\partial_\alpha\gamma^\alpha + \kappa(1+\tilde{a})\right]\varphi_b = 0 \quad , \tag{86}$$

where

$$\Phi' \equiv S\Phi \equiv \begin{bmatrix}\Phi_a{}' \\ \Phi_b{}'\end{bmatrix} = \begin{bmatrix}\varphi_a \\ \varphi_b\end{bmatrix} \quad . \tag{87}$$

Eq.(43) holds too.Thus,analogously to (45) we have

$$\Phi' = \begin{bmatrix}\varphi_a \\ N_b\varphi_a{}^*\end{bmatrix} \qquad \varphi_a = \begin{bmatrix}\varphi_{a1} \\ \varphi_{a2} \\ \varphi_{a3} \\ \varphi_{a4}\end{bmatrix} \tag{88}$$

and

$$\Phi = \frac{1}{\sqrt{2}}\begin{bmatrix} -\varphi_{a1} - \varphi_{a1}{}^* \\ -i\varphi_{a1} + i\varphi_{a1}{}^* \\ \varphi_{a2} + \varphi_{a2}{}^* \\ i\varphi_{a2} - i\varphi_{a2}{}^* \\ i\varphi_{a3} - i\varphi_{a3}{}^* \\ -\varphi_{a3} - \varphi_{a3}{}^* \\ -i\varphi_{a4} + i\varphi_{a4}{}^* \\ -\varphi_{a4} - \varphi_{a4}{}^* \end{bmatrix} = \sqrt{2}\begin{bmatrix} -\mathrm{Re}\,\varphi_{a1} \\ \mathrm{Im}\,\varphi_{a1} \\ \mathrm{Re}\,\varphi_{a2} \\ -\mathrm{Im}\,\varphi_{a2} \\ -\mathrm{Im}\,\varphi_{a3} \\ -\mathrm{Re}\,\varphi_{a3} \\ \mathrm{Im}\,\varphi_{a4} \\ -\mathrm{Re}\,\varphi_{a4} \end{bmatrix} \tag{89}$$

The equation for $A^\alpha$ comes from $\delta_{A^\alpha}\int L \,\frac{1}{c} d^4x = 0$ and is given by

$$\partial_\mu\partial^\mu A^\alpha = 4\pi e\left[\kappa^2\overline{\Phi}\eta^\alpha\Phi + \frac{\overline{\Psi}\eta^\alpha\Psi}{1-a^2} - \frac{2\frac{e}{K}A^\alpha\overline{\Psi}(1-a)\Psi}{(1-a^2)^2}\right] + \left(\frac{4\pi}{c}j_{ext}\right) \quad . \tag{90}$$

For solutions (83) it becomes

$$\partial_\mu\partial^\mu A^\alpha = 4\pi e \cdot 2\kappa^2\overline{\Phi}\eta^\alpha\Phi + \left(\frac{4\pi}{c}j^\alpha_{ext}\right) \quad . \tag{91}$$

The equations (84) and (91) determine behaviour of the system.As one can see it contains the self-interaction too.It is like in the classical theory.Disregarding the self-interaction (like in the classical theory) one obtains Dirac equation in the external field.The problem of the self-interaction we consider elsewhere (see also [8-9]).

Constants of motion

The scalar constant of motion comes from

$$\partial_\alpha \overline{\Psi}_I \eta^\alpha \Psi_I = 0 \tag{92}$$

and is given by

$$Q = \text{const.}_Q \int \Psi_I^+ \Psi_I d^3x \quad . \tag{93}$$

For the solutions (83) it is

$$Q = \text{const.}_Q \int 2\kappa^2 \Phi^+ \Phi d^3x \quad . \tag{94}$$

Energy-momentum and spin projection $S^3$ are determined by

$$P^\alpha = \text{const.}_P \frac{K}{2\kappa} \int \left[ \Psi_I^+ i\partial^\alpha \Psi_I + \left( \Psi_I^+ i\partial^\alpha \Psi_I \right)^* \right] d^3x \quad , \tag{95}$$

$$S^3 = \text{const.}_M \frac{K}{4\kappa} \int \left[ \Psi_I^+ \eta^1 \eta^2 i \Psi_I + \left( \Psi_I^+ \eta^1 \eta^2 i \Psi_I \right)^* \right] d^3x \quad . \tag{96}$$

For the solutions (83) they are

$$P^\alpha = \text{const.}_P (-2\kappa K) \int \Phi^+ \partial^\alpha N \Phi d^3x \quad , \tag{97}$$

$$S^3 = \text{const.}_M (\kappa K) \int \Phi^+ \eta^1 \eta^2 \Phi d^3x \quad . \tag{98}$$

<u>Example</u>

$$A^\alpha \longrightarrow \left( \frac{|e|}{r}, 0 \right) \quad , \text{(hydrogen atom without self-interaction)} \quad ,$$

$$\varphi_a = \begin{bmatrix} g_{0-1} \begin{bmatrix} y_{oo} \\ 0 \end{bmatrix} \\ if_{0-1} \begin{bmatrix} -\sqrt{\frac{1}{3}} y_{10} \\ \sqrt{\frac{2}{3}} y_{11} \end{bmatrix} \end{bmatrix} e^{-ik_o x^o} \quad \text{(Dirac ground state)} ,$$

$$\Phi = \frac{1}{\sqrt{2\pi}} \begin{bmatrix} -g_{o-1}(r) \cos k_o x^o \\ -g_{o-1}(r) \sin k_o x^o \\ 0 \\ 0 \\ f_{o-1}(r) \cos\vartheta \cos k_o x^o \\ f_{o-1}(r) \cos\vartheta \sin k_o x^o \\ f_{o-1}(r) \sin\vartheta (\cos\varphi \cos k_o x^o + \sin\varphi \sin k_o x^o) \\ f_{o-1}(r) \sin\vartheta (\sin\varphi \cos k_o x^o - \cos\varphi \sin k_o x^o) \end{bmatrix}$$

The corresponding scalar and energy constant of motion are given by

$$Q = const._Q \frac{\kappa^2}{\pi} \int \left( g_{o-1}^2 + f_{o-1}^2 \right) d^3x \quad ,$$

$$P^o = const._P \left( \frac{\kappa K}{\pi} \right) k_o \int \left( g_{o-1}^2 + f_{o-1}^2 \right) d^3x \quad .$$

Selection

$$const._Q \frac{\kappa^2}{\pi} \int \left( g_{o-1}^2 + f_{o-1}^2 \right) d^3x = 1 \quad , \quad const._P \frac{\kappa^2}{\pi} \int \left( g_{o-1}^2 + f_{o-1}^2 \right) d^3x = \frac{1}{c} \quad ,$$

yields

$$Q = 1,$$
$$P^o = \hbar k_o,$$

in agreement with the conventional quantum theory,quantitatively as well as qualitatively.But the field $\Phi$ is analogous to the electromagnetic field,only it is constructed from spinor waves.

## 4.COMMENTS

The presented results show that the Dirac field is a classical field similar to the electromagnetic field but constructed from spinors.They also put some light on context and origin of the quantum physics.It tells us more about the nature of physical reality.

References

[1] J.Brana and K.Ljolje , FIZIKA (Zagreb) 10 (1978)85

[2] K.Ljolje and S.Vobornik , FIZIKA (Zagreb) 12 (1979)171

[3] J.Brana and K.Ljolje , FIZIKA (Zagreb) 12 (1980)287

[4] K.Ljolje , Forthschr. Phys. 36 (1988) 1, 9

[5] K.Ljolje , Problems in Quantum Physics II , Gdańsk '89 (1989) 393

[6] J.Brana and K.Ljolje , Akademija nauka i umjetnosti Bosne i Hercegovine , RADOVI-LIX (1976) 5

[7]K.Ljolje , A Real Spinor Field , supported by Soros Foundation , Sarajevo (1996)

[8] F.Rohrlich , Classical Charged Particles , Addison-Wesley , Reading , 1964
[9] S.Botrić and K.Ljolje , Il Nuovo Cimento 107B , N.1 (1992) 51